\documentclass[final,5p,times,twocolumn]{elsarticle}
\usepackage[T1]{fontenc}
\usepackage{hyperref}

\usepackage{amssymb}

\usepackage{lineno}

\journal{Nuclear Instruments \& Methods in Physics Research, Section A}

\begin{document}

\begin{frontmatter}



\title{Current status and operation of the H.E.S.S. array of imaging atmospheric Cherenkov telescopes}

\author[1]{Stefan Ohm}
\ead{stefan.ohm@desy.de}
\author[2]{Stefan Wagner}
\author{for the H.E.S.S. Collaboration}
\ead{contact.hess@hess-experiment.eu}

\affiliation[1]{organization={Deutsches Elektronen-Synchrotron DESY},
            addressline={Platanenallee 6}, 
            city={Zeuthen},
            postcode={15738}, 
            country={Germany}}

\affiliation[2]{organization={Landessternwarte, Universit\"at Heidelberg},
            addressline={K\"onigstuhl}, 
            city={Heidelberg},
            postcode={69117}, 
            country={Germany}}

\begin{abstract}
The High Energy Stereoscopic System (H.E.S.S.) is an array of five imaging atmospheric Cherenkov telescopes (IACTs) to study gamma-ray emission from astrophysical objects in the Southern hemisphere. It is the only hybrid array of IACTs, composed of telescopes with different collection area and footprint, individually optimised for a specific energy range. Collectively, the array is most sensitive to gamma rays in the range of 100\,GeV to 100\,TeV. The array has been in operation since 2002 and has been upgraded with new telescopes and cameras multiple times. Recent hardware upgrades and changes in the operational procedures increased the amount of observing time, which is of key importance for time-domain science. H.E.S.S. operations saw record data taking in 2020 and 2021 and we describe the current operations with specific emphasis on system performance, operational processes and workflows, quality control, and (near) real-time extraction of science results. In light of this, we will briefly discuss the early detection of gamma-ray emission from the recurrent nova RS Oph and alert distribution to the astrophysics community.

\end{abstract}


\end{frontmatter}


\section{Introduction}
\label{intro}
The High Energy Stereoscopic System (H.E.S.S.) is an array of five imaging atmospheric Cherenkov telescopes (IACTs) operating in the Khomas Highland in Namibia. In October 2019, the H.E.S.S. collaboration entered a first extension phase that lasted until the end of September 2022. The main operations goals for this extension phase were an increase in telescope and instrumental reliability as well as a significant increase in total observing time. This contribution describes the main technical activities and changes in operation procedures that were implemented in the first extension phase. They led to a $\sim$50\% increase in yearly total observation time as well as a significant improvement in telescope up-time, reaching 98\% per telescope, and overall data quality. The three main elements to achieve these goals are: 1) Hardware upgrades and maintenance; 2) Observations under moderate moonlight and twilight; 3) Observation procedures and data quality monitoring.

\section{Hardware upgrades and availability}
\label{hardware}
Ever since the installation of the first telescope in 2002, the H.E.S.S. system underwent frequent hardware upgrades and maintenance efforts to improve or maintain telescope performance. 
\begin{figure}[t!]
    \centering
    \includegraphics[width=0.475\textwidth]{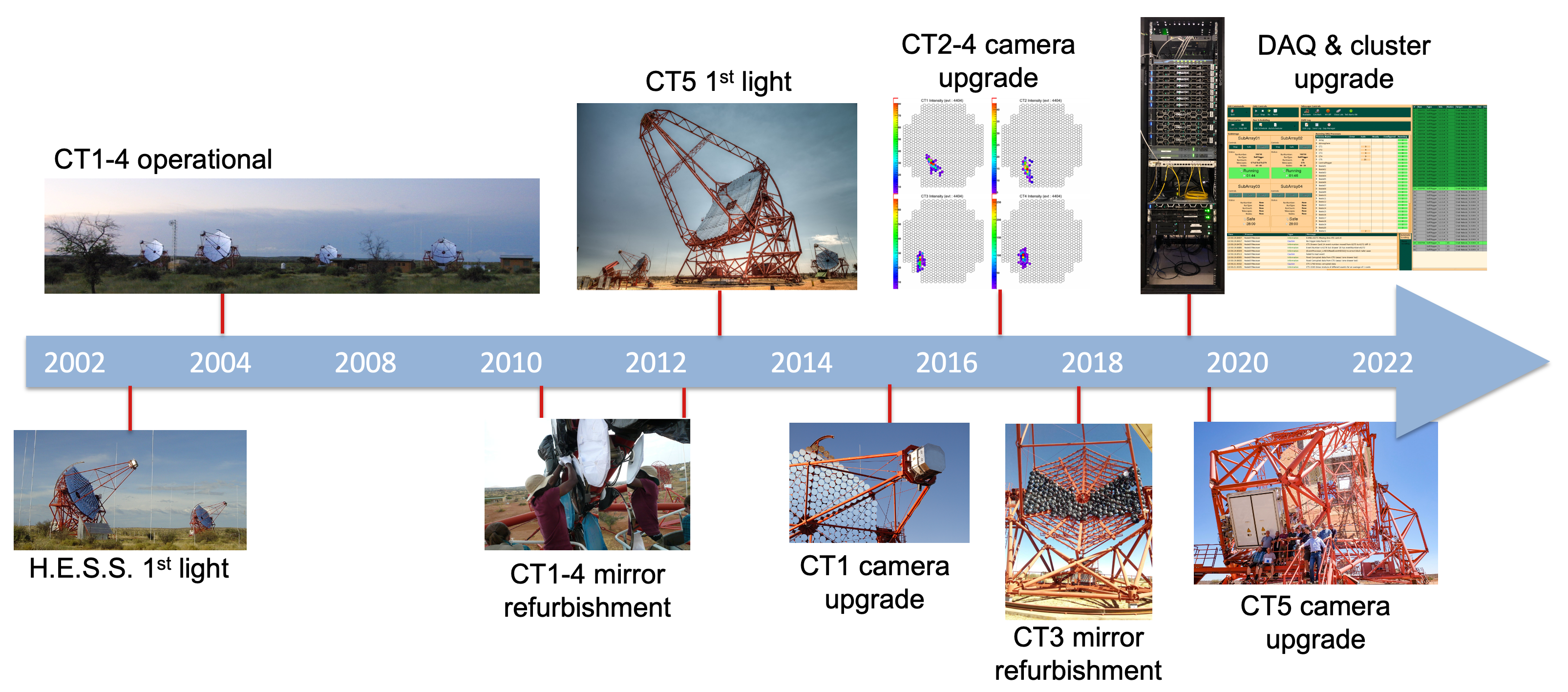}
    \caption{Timeline illustrating major H.E.S.S. hardware upgrades and maintenance campaigns over the last 20 years. Additionally, regular Winston cone exchanges and cleanings, Cherenkov camera gain adjustment, and other works have been conducted but are not listed here.}
    \label{fig:hardware}
\end{figure}
Figure~\ref{fig:hardware} shows a timeline with major hardware upgrades conducted throughout the 20-year history of H.E.S.S. The most elaborate upgrade was the installation of the large CT~5 telescope in 2012, making H.E.S.S. the only operational hybrid IACT system to date. In 2015/2016 the four Cherenkov cameras of the smaller H.E.S.S. telescopes were upgraded (HESS-IU)~\cite{HESS-IU}. At the beginning of the first extension phase, the camera of the CT~5 telescope was replaced with a Cherenkov Telescope Array (CTA) prototype FlashCam camera~\cite{FC:concept}. The installation and integration of the camera went very smoothly with detection of the Crab Nebula on the first night of observations~\cite{FC:scienceverification}. Since installation end of 2019, FlashCam operation is very stable with an up-time exceeding 99\% and fulfilling CTA requirements~\cite{FC:performance}. In preparation for the upcoming CT~5 camera upgrade, and expected longer-term H.E.S.S. operation, the data acquisition system (DAQ) as well as the on-site computing cluster underwent a major upgrade in early 2019. This upgrade was finalised within 3 months after the integration of the new CT~5 camera in the array~\cite{Zhu2021_DAQCluster}. The overall downtime due to DAQ problems could be halved to $\sim$0.5\% in the first extension phase. The increased hardware availability of the cameras in particular led to a more stable overall system and reduced downtime of other components as well. With further improvements made to other sub-systems like the HESS-IU cameras, or the tracking control system, we estimate the improvement in additional observation time as $100-150$ hours per year. 
\begin{figure*}[t!]
    \centering
    \includegraphics[width=0.975\textwidth]{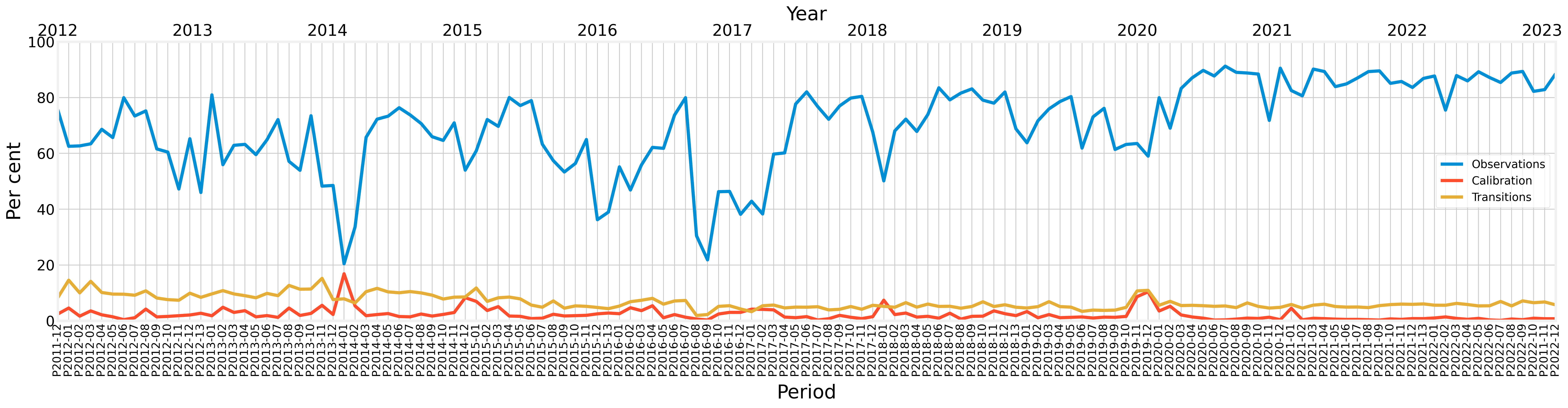}
    \caption{H.E.S.S. data taking efficiency weighted by the fraction of telescopes participating in observations and corrected for bad weather. Some of the strong dips in efficiency are related to upgrades and maintenance campaigns as indicated in e.g. Figure~\ref{fig:hardware}.}
    \label{fig:efficiencies}
\end{figure*}
Figure~\ref{fig:efficiencies} shows the long-term data-taking efficiency since the installation of CT~5 in 2012. A significant increase in average efficiency (on-target data taking, weighted by participating telescopes, excluding bad weather) from $\sim$($60-80$)\% before October 2019 to $\sim$90\% after is clearly visible. Around ($5-7$)\% of the total available observation time is spent on telescope slewing between different targets in the sky.

\section{Observations under moderate moonlight and twilight}
\label{moonlight}
Another major goal for the first H.E.S.S. extension phase was to increase total observation time by extending routine observations into periods of moderate amounts of moonlight. Observations with imaging atmospheric Cerenkov telescopes are conducted with very sensitive detectors that were traditionally only operated under dark-sky conditions. Initial tests of observations under moonlight were conducted in 2019 and motivated by the discovery of GRB\,190114C with the MAGIC telescopes in moonlight observations~\citep{MAGIC:190114C}. The camera hardware settings were fixed in March 2020 and the system basically prepared for continuous moonlight observations. To increase the lifetime of the HESS-IU cameras, one single gain setting was defined for both, observation in astronomical darkness as well as under moderate moonlight~\citep{HESS-IU:gain}. The full implementation of regular moonlight observations, including adaptations in e.g. the scheduling, or transient follow-up system was finalised in January 2021. Moonlight observations are conducted up to a moon phase of 40\%, a target-to-moon separation between 45$^\circ$ and 145$^\circ$, and a maximum predicted night-sky-background (NSB) level of 3.5 times the dark NSB of 100\,MHz per pixel in the HESS-IU cameras. Observations with the H.E.S.S. array can now be conducted for around 250\,hours extra each year without a significant loss in sensitivity or performance during periods of moderate moonlight.

\begin{figure}[ht!]
    \centering
    \includegraphics[width=0.475\textwidth]{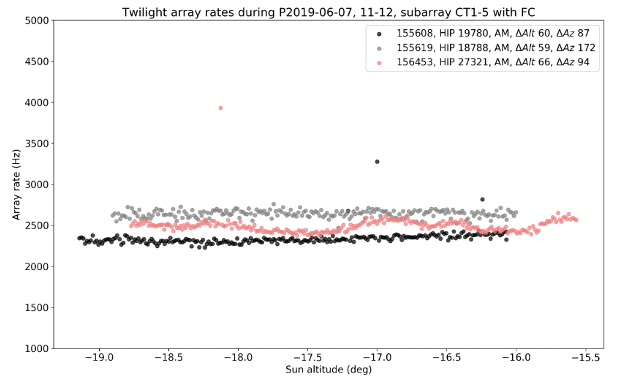}
    \caption{CT1-5 array trigger rates as a function of sun altitude for three different telescope pointing directions. No significant increase in trigger rate is seen when going from astronomical darkness to astronomical twilight at $-16^\circ$.}
    \label{fig:twilight}
\end{figure}

A further increase in available observation time resulted from a widening of the observing window with an earlier start and a later end of observations each night. Historically, H.E.S.S. observations were conducted in astronomical darkness, starting and ending observations at sun elevation angles of $-18^\circ$. Careful testing of the system behaviour was performed for observations under astronomical twilight in mid and end of 2019. Figure~\ref{fig:twilight} shows the CT1-5 array trigger rates as a function of sun elevation angle for different telescope pointing directions with respect to the sun position. Only a slow increase in trigger rates can be seen with increasing sun elevation angles, which confirms that it is safe to operate the telescopes in astronomical twilight. The new setting of $-16^\circ$ for the maximum sun elevation angle has been implemented in January 2021 and results in an additional observation time of about 70\,hours per year for the above moonlight settings.

\section{Observation procedures and data quality monitoring}
\begin{figure}[ht!]
    \centering
    \includegraphics[width=0.475\textwidth]{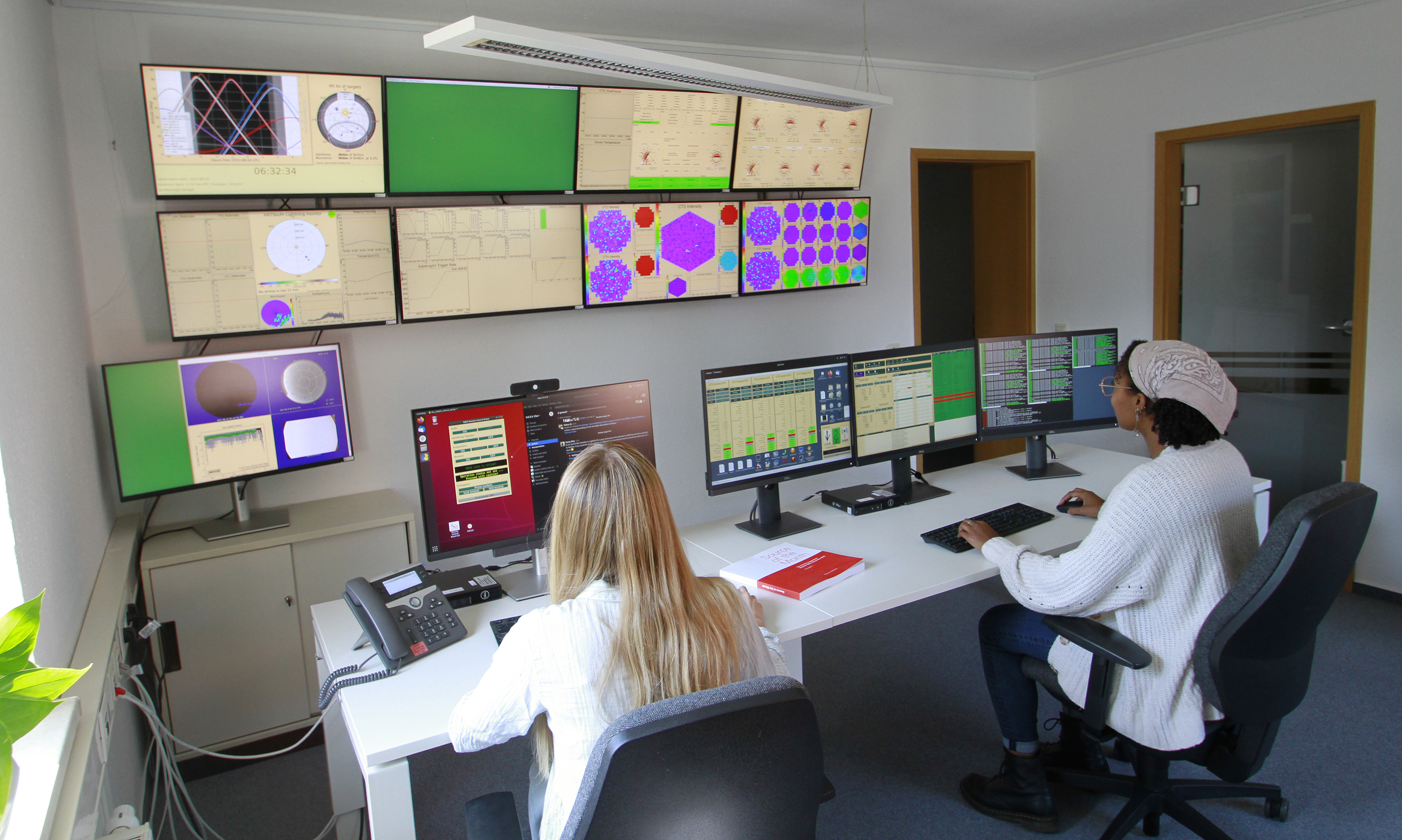}
    \caption{H.E.S.S. remote control room established at DESY Zeuthen. The setup closely resembles the control room on the H.E.S.S. site and allows for joint as well as stand-alone operation of the telescopes.}
    \label{fig:remote}
\end{figure}

With the outbreak of the Covid-19 pandemic, a significant change in the H.E.S.S. observation procedure had to be implemented. Before Corona, two on-site experts trained a team of off-site personnel from H.E.S.S. partner institutes that typically conducted one observing shift between two full-moon periods. While travel was severely impacted after March 2020, observations were performed by on-site shift experts, supported by newly hired local shifters, as well as students and members of the University of Namibia, who were still allowed to travel to/from the site. This strong support and change in shift operation allowed H.E.S.S. to conduct science observations throughout the entire pandemic. Since travel restrictions have been eased, H.E.S.S. is operating in a mode where at least one professional local shifter conducts observations throughout an observing shift and is supported by shifters from H.E.S.S. partner institutes. This mode of operation guarantees reliable and stable operation through experienced local shift personnel while being able to train junior researchers from abroad in the operation of the H.E.S.S. telescopes. Furthermore, a remote H.E.S.S. operations room has been established at DESY in Zeuthen and is shown in Fig.~\ref{fig:remote}\footnote{The remote control room was featured in the H.E.S.S. Source-of-the-Month of \href{https://www.mpi-hd.mpg.de/hfm/HESS/pages/home/som/2022/10/}{October 2022}.}.
The remote control room increases flexibility in telescope operation, reduces CO$_2$ footprint stemming from international travel, as well as allows for training of technical experts before e.g. conducting upgrade or maintenance campaigns on site. The remote control room has been successfully used for all these purposes and further remote control rooms are currently being set up at other H.E.S.S. member institutes.

Another major effort went into the documentation of telescope operation, a compilation of How-To's, and troubleshooting guidelines for system or sub-system errors and failures. This activity was particularly important given the changes in operational procedures and extended observations under moonlight and twilight. In particular, the problem troubleshooting guidelines are continuously updated and assure that known problems are identified and resolved as fast as possible during data taking. A newly established off-site data quality team, which rotates and typically consists of two \textit{day shifters} per observing period monitors the quality of the data taken in the previous night. Data quality from the lowest (e.g. camera, trigger) to the highest (shower images, real-time-analysis sky maps) are checked and errors are flagged. The data quality team also monitors the long-term (sub-)system behaviour such as the muon efficiency, or the number of broken/deactivated pixels in the Cherenkov cameras. Problems encountered during the night and discovered the next day are documented in \textit{Shift Workbooks} that serve as a central hub to collect the monthly shift schedule as well as target-of-opportunity (ToO) observations. Discussions between on-site and remote shift crew, day shifters, and sub-system experts are mainly conducted via Slack messenger. Weekly virtual meetings between sub-system experts and the shift crews summarise data taking and provide a forum for more detailed discussions. Monthly \textit{Operations calls} are held in between observation periods and discuss longer-term operational activities like the implementation of new observing modes or prepare and inform about maintenance campaigns. These platforms have been found to be critical for information exchange and troubleshooting of issues. We estimate that through the implementation of these procedures and additional 50\,hours of observations per year could be achieved. A revision of the calibration strategy resulted in another 15\,hours of extra observation time per year. Monthly summary mails about technical activities on- and off-site, telescope operations, and data-taking efficiency are prepared and sent to the collaboration.

\section{H.E.S.S. operation and optimisation for transient science}

\begin{figure*}[t!]
    \centering
    \includegraphics[width=0.88\textwidth]{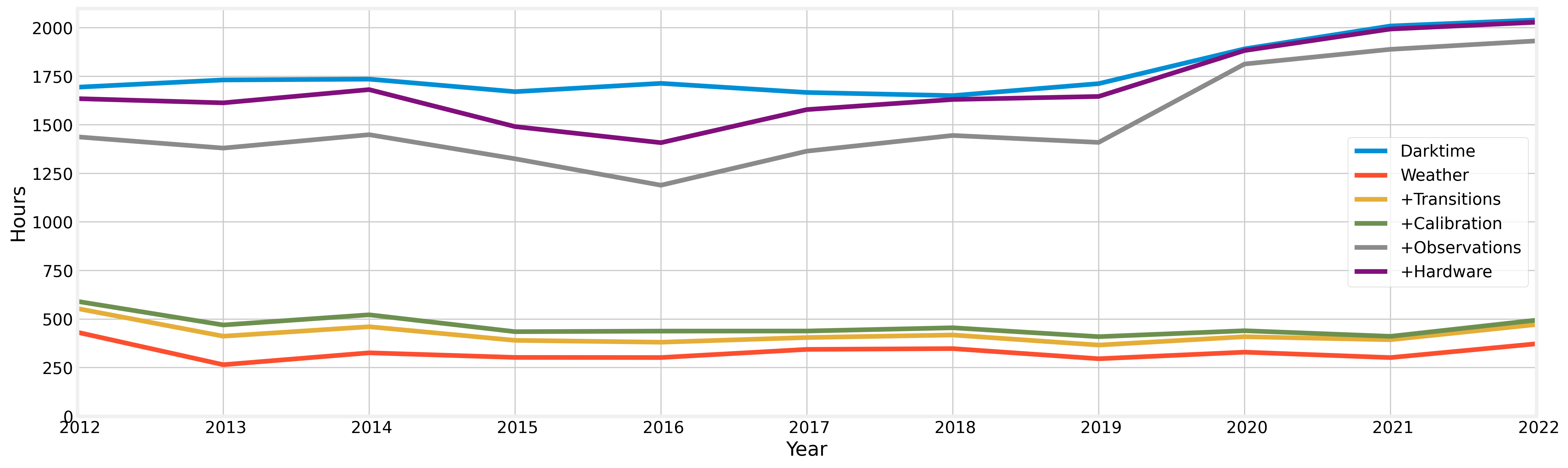}
    \includegraphics[width=0.88\textwidth]{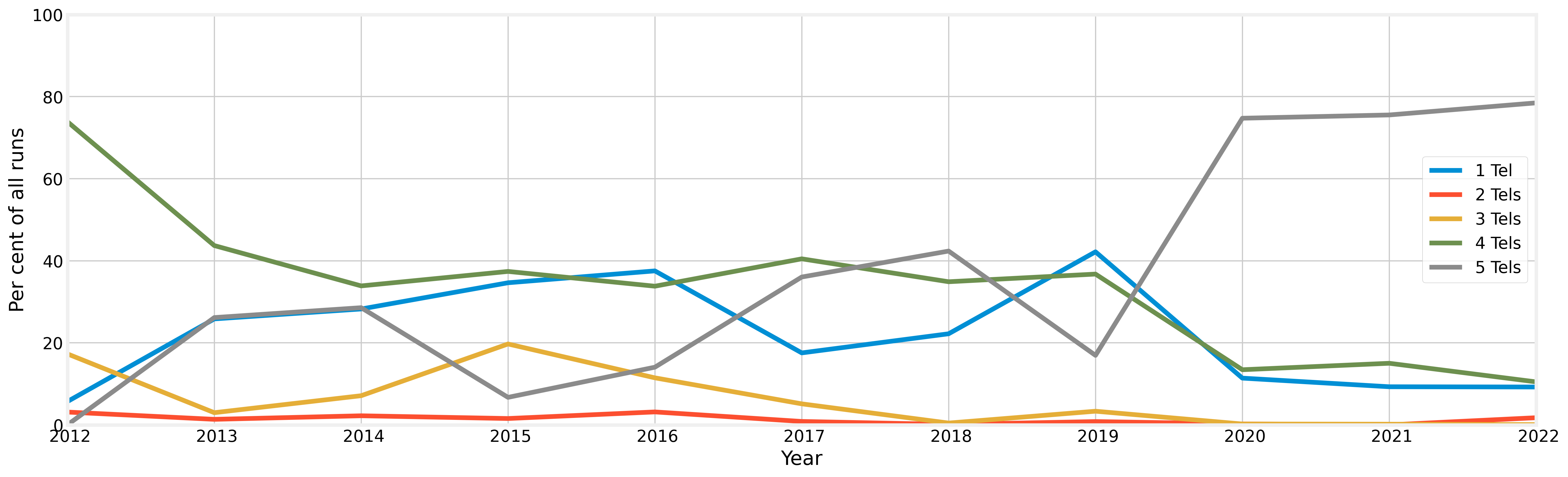}
    \caption{Top: Time available for science observations per year (light blue). Also shown are the contributions for various telescope operation modes and downtimes. The area between the grey (Observations) and purple (Hardware problems) curve is reduced since 2019 and stable throughout the first extension. Bottom: Since 2020, most observations are conducted with all 5 telescopes. The remainder of 1 and 4 telescope runs are comprised of camera warm-up and calibration runs.}
    \label{fig:totals}
\end{figure*}

The efforts described above resulted in a significant increase in operational efficiency and record-breaking data-taking in 2020 and 2021. In particular, 2021 saw a total observation time exceeding 1500\,hours (or 17\% duty cycle) also thanks to favourable weather conditions. That 2021 was not an exception can also be seen in Fig.~\ref{fig:totals}. Compared to previous years more than 300 hours of extra time are now available for science observations. While downtime due to weather and transitions can hardly be reduced, the downtime due to hardware problems was reduced considerably to the few percent level. Another advantage of the much more stable operation is the homogeneity of the data that is taken. Since 2020, the vast majority of data is taken with the full 5-telescope array (cf. Fig.~\ref{fig:totals}).
Throughout the first extension phase, H.E.S.S. operations have been optimised to maximise telescope availability for ToO observations. Continued improvements have been made to the H.E.S.S. transients alert system, real-time-analysis, and next-day on-site analysis capabilities~\citep{HESS:TransientsSystem} as well as data transfer off-site for final calibration and data analysis. All these activities have ultimately allowed H.E.S.S. to reduce the time between data taking and final analysis to $<$2 days. Furthermore, continued improvements in the atmosphere monitoring and treatment of observation conditions in Monte Carlo simulations and instrument response functions are being implemented. The discovery of the first Galactic transient in very high energy gamma rays, the recurrent nova RS~Oph was the culmination of all these activities \citep{HESS:RSOph}. Data were taken during moon time, astronomical twilight, and under strongly varying atmospheric conditions (which were corrected for in the final analysis). Furthermore, the fast real-time and next-day analysis informed shifters and experts early about the detection and source properties, which allowed H.E.S.S. to inform the community via Astronomers Telegrams \citep{rsoph_detection_atel, rsoph_spectrum_atel}. This is just an example demonstrating how the developments implemented in the first H.E.S.S. extension phase and before put the experiment in an ideal role for time-domain multi-messenger and multi-wavelength astronomy.

H.E.S.S. and its various sub-systems are undergoing a phase to prepare a low-maintenance mode in which no further major upgrades to hardware components or changes in settings are envisaged. This is aimed at minimising the maintenance effort and provide stability in data-taking efficiency during the transition period towards CTA installation and buildup. Recognition of the (mostly) junior researchers working on technical tasks and maintaining the various sub-systems is elevated through newly established public and citeable internal notes that are published via Zenodo, and on the official H.E.S.S. webpages, including appropriate advertisement through the various social media channels. Lessons learnt are communicated through these notes internally and to the community to maximise knowledge transfer from H.E.S.S. as the only hybrid IACT system to CTA.

\section{Acknowledgement}
Full H.E.S.S. acknowledgements can be found \href{https://www.mpi-hd.mpg.de/hfm/HESS/pages/publications/auxiliary/HESS-Acknowledgements-2021.html}{here}.





\bibliographystyle{elsarticle-num}
\bibliography{HESS_Operations_RICH2022}
\end{document}